\documentclass[aps,prl,twocolumn,showpacs,amsmath,amssymb,floatfix]{revtex4}
\usepackage{bm}
\usepackage{bbm}

\usepackage{graphicx, pictex}

\begin{document}

\title{Ordered states of adatoms on graphene.}
\author{ Vadim~V.~Cheianov$^1$, Olav Sylju\aa sen$^2$, B.L. Altshuler$^{1,3}$%
, and~Vladimir~Fal'ko$^1$}
\affiliation{$^1$ {Physics Department, Lancaster University, Lancaster LA1 4YB, UK } \\
$^2${Institute for Physics, University of Oslo, 1048 Blindern N-0316 Oslo,
Norway}\\
$^3${Physics Department, Columbia University, 538 West 120th Street, New
York, NY 10027, USA} }

\begin{abstract}
We show that a dilute ensemble of epoxy-bonded adatoms on graphene has a
tendency to form a spatially correlated state accompanied by a gap
in graphene's electron spectrum. This effect emerges from the
electron-mediated interaction between adatoms with a peculiar $1/r^{3}$
distance dependence. The partial ordering transition is described by a
random bond three state Potts model.
\end{abstract}

\maketitle

Graphene (monolayer of graphite) is a truly two-dimensional crystal, just
one-atom-thick \cite{Geim1}. It is a gapless semiconductor with charge
carriers mimicking relativistic dynamics of massless Dirac fermions \cite{Review},
a peculiarity dictated by the bonding of carbon atoms into a highly symmetric
honeycomb lattice. Graphene  can host various adsorbents, in particular atoms,
retaining its own structural integrity.  Such chemisorbed atoms
(adatoms) may strongly affect electronic properties of graphene
\cite{Geim3,Fuhrer,Lanzara,Geim2} introducing symmetry-breaking perturbations
into the lattice. The type of symmetry breaking depends on the position of
the adatom in the hexagonal unit cell of the crystal. In particular, alkali
atoms position themselves over the centres of the hexagons \cite{K}.
Oxygen, nitrogen, boron, or an additional carbon \cite{C-on-G} prefer
'epoxy' bonded positions (e-type) and reside above the middle of a
carbon-carbon bond. Atomic hydrogen and halogens reside in the
symmetric on-site position above the carbon (s-type) \cite{H-on-G}. It has
also been noticed that a pair of hydrogen atoms on the neighbouring sites of
graphene lattice forms a stable
H-H dimer which acts as an e-type adsorbent \cite{H2-on-G}.

\begin{figure}
\includegraphics[width= 0.48 \textwidth]{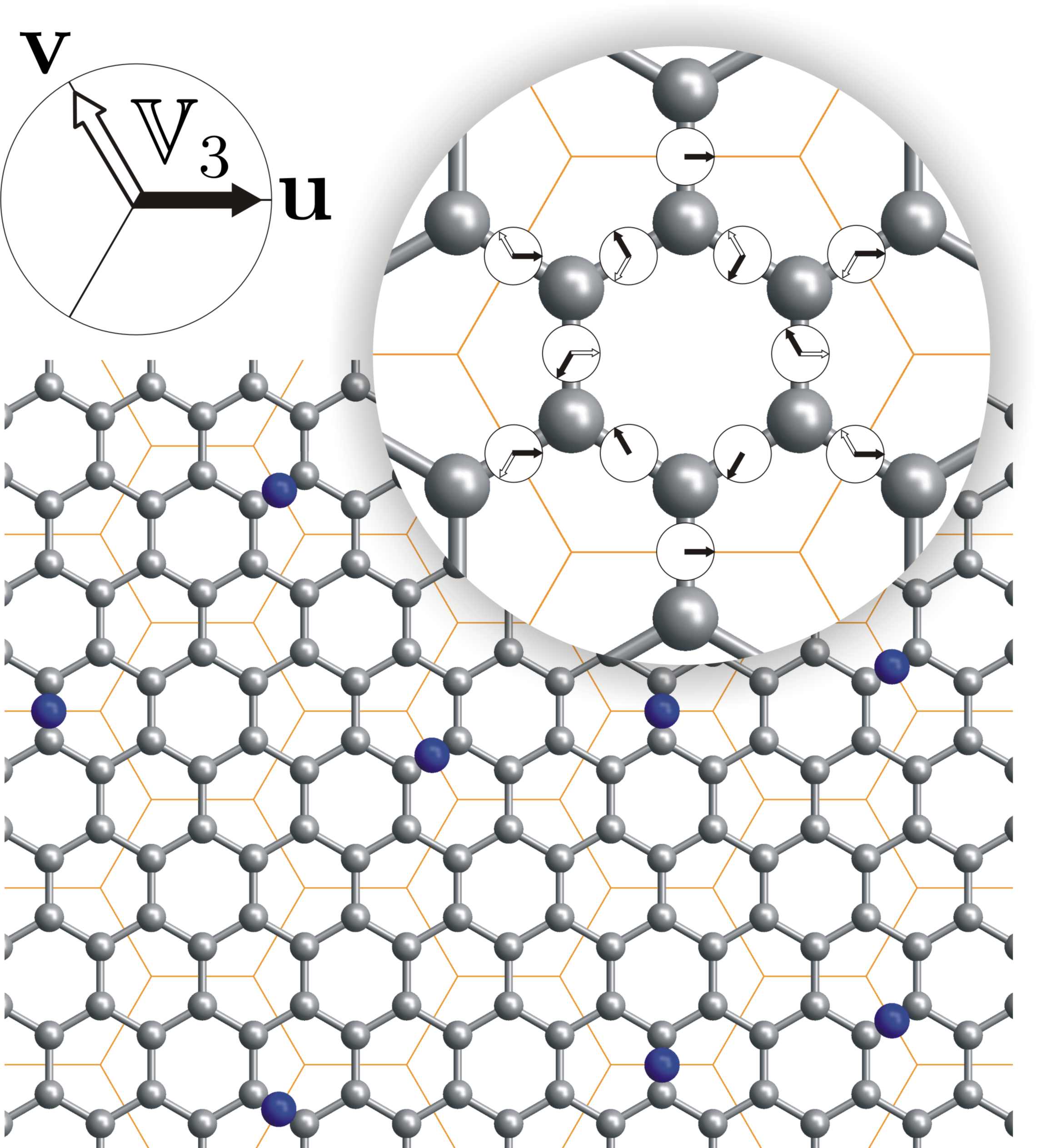}
\caption{An ordered ensemble of adatoms (blue) on graphene lattice:
all impurity atoms occupy positions at the intersection of graphene
bonds and the bonds of the underlying fictitious superlattice. The inset
shows how the superlattice originates from the scattering matrix $\hat{W},$
Eq.~\eqref{lambdaT}. To each bond there corresponds a
pair of vectors $(\mathbf{u},\mathbf{v})$ forming a periodic pattern with
three times graphene lattice's period. }
\label{Fig01}
\end{figure}

Here we predict that an ensemble of e-type adatoms (those perturbing
C-C bonds) tend to order, mimicking a superlattice structure, even when graphene
coverage by adsorbents is low. The underlying mechanism is a
long-range electron-mediated interaction between adatoms similar to the
RKKY exchange between  localized spins
in metals \cite{RKKY1954}. The effect is peculiar to graphene.
Unlike metals, charge neutral graphene
has a point-like Fermi surface
positioned in the corners $\mathbf{K}$ and $\mathbf{K}^{\prime }=-\mathbf{K}$
of the hexagonal Brillouin zone -- called valleys. The electron density
of states vanishes at the Fermi level.
As a result, the Friedel oscillations in charge neutral
graphene are commensurate with its honeycomb lattice and decay as the inverse
cube of the distance to the adatom \cite{Fertig}. We show that such an
interaction in a dilute ensemble of e-type adsorbents may result in
their partial ordering associated with a superlattice structure with the
unit cell three times larger than in graphene, as illustrated
in Fig.~\ref{Fig01}.
We present our results in the following order.
Starting with a particular tight-binding model for an e-type adsorbent,
we determine the form of a perturbation it creates for the
electrons in graphene. Using group theory we classify such interactions
beyond a specific microscopic model and determine the conditions under
which RKKY interaction between adatoms leads to a partially ordered state
with a gapful electronic spectrum. We conclude by discussing experimental
signatures of the effect.

\begin{table*}[tbp]
\caption{Symmetry-based classification of RKKY interactions for different
types of adatoms.
The parameterization of the adatom
scattering matrix $\hat W_{a}$ is given in terms of an orbit $\mathcal W_i$
of the lattice symmetry group in a given irrep $i.$
The potential $\Omega _{ab}^{(i)},$ is a contribution of the given orbit to
the adatom-adatom interaction, Eqs.~\eqref{Omint}. The last column indicates
ordering favored by the given interaction term. }
\label{tab:invars}
\begin{tabular}{|l|l|l|l|p{1.8 cm}|}
\hline
Irrep $i$ & Orbit $\mathcal{W}_{i}$ & $\Omega _{ab}^{(i)}$ & type: position
& Partial order
\\ \hline\hline
$A_{1}$ & $\mathbbm1$ & $-1/2$ & All & ---
\\ \hline
$B_{2}$ & $s\Sigma _{z}\Lambda _{z}$, $s=\pm 1$ & $s_{a}s_{b}$ & s:%
\mbox{
\beginpicture

\setcoordinatesystem units <0.08 pt,0.08 pt> \setplotarea x from -130 to 130, y from -180 to 180

\plotsymbolspacing=.2pt
\linethickness=2 pt
\setplotsymbol ({.})
\setlinear
\plot 0 100
      87 50 /
\plot  87 50
      87 -50 /
\plot  87 -50
       0 -100 /
\plot  0 -100
       -87 -50 /
\plot  -87 -50
       -87 50 /
\plot  -87 50
       0 100 /
\setsolid
\plot 0 100
      0 150 /
\plot  87 50
      130 75 /
\plot  87 -50
       130 -75 /
\plot  0 -100
       0 -150 /
\plot  -87 -50
       -130 -75 /
\plot  -87 50
       -130 75 /

 \put {\Large $\bullet$} at 0 90
\endpicture} & Sublattice
 \\ \hline
$E_{1}^{\prime }$ & $\Sigma _{z}\mathbf{(}\mathbf{\Lambda }\cdot \mathbf{u})$%
, $\mathbf{u}\in \mathbb{V}_{3}$ & $\mathbf{u}_{a}\cdot \mathbf{u}_{b}$ & e:%
\mbox{
\beginpicture

\setcoordinatesystem units <0.08 pt,0.08 pt> \setplotarea x from -130 to 130, y from -180 to 180

\plotsymbolspacing=.2pt
\linethickness=2 pt
\setplotsymbol ({.})
\setlinear
\plot 0 100
      87 50 /
\plot  87 50
      87 -50 /
\plot  87 -50
       0 -100 /
\plot  0 -100
       -87 -50 /
\plot  -87 -50
       -87 50 /
\plot  -87 50
       0 100 /
\setsolid
\plot 0 100
      0 150 /
\plot  87 50
      130 75 /
\plot  87 -50
       130 -75 /
\plot  0 -100
       0 -150 /
\plot  -87 -50
       -130 -75 /
\plot  -87 50
       -130 75 /

 \put {\Large $\bullet$} at -87 -03

\endpicture}
\mbox{
\beginpicture
\setcoordinatesystem units <0.08 pt,0.08 pt> \setplotarea x from -130 to 130, y from -180 to 180
\plotsymbolspacing=.2pt
\linethickness=2 pt
\setplotsymbol ({.})
\setlinear
\plot 0 100
      87 50 /
\plot  87 50
      87 -50 /
\plot  87 -50
       0 -100 /
\plot  0 -100
       -87 -50 /
\plot  -87 -50
       -87 50 /
\plot  -87 50
       0 100 /
\setsolid
\plot 0 100
      0 150 /
\plot  87 50
      130 75 /
\plot  87 -50
       130 -75 /
\plot  0 -100
       0 -150 /
\plot  -87 -50
       -130 -75 /
\plot  -87 50
       -130 75 /

 \put {\Large $\bullet$} at -87 -53
 \put {\Large $\bullet$} at -87  53

\endpicture} & $3\times$ unit cell superlattice
 \\ \hline
$E_{2}$ & $\Lambda _{z}\mathbf{(}\mathbf{\Sigma }\cdot \mathbf{v})$, $%
\mathbf{v}\in \mathbb{V}_{3}$ & $\mathbf{v}_{a}\cdot \mathbf{v}_{b}-\frac{3}{%
2}(\mathbf{n}\cdot \mathbf{v}_{a})(\mathbf{n}\cdot \mathbf{v}_{b})$ & e:%
\mbox{
\beginpicture

\setcoordinatesystem units <0.08 pt,0.08 pt> \setplotarea x from -130 to 130, y from -180 to 180

\plotsymbolspacing=.2pt
\linethickness=2 pt
\setplotsymbol ({.})
\setlinear
\plot 0 100
      87 50 /
\plot  87 50
      87 -50 /
\plot  87 -50
       0 -100 /
\plot  0 -100
       -87 -50 /
\plot  -87 -50
       -87 50 /
\plot  -87 50
       0 100 /
\setsolid
\plot 0 100
      0 150 /
\plot  87 50
      130 75 /
\plot  87 -50
       130 -75 /
\plot  0 -100
       0 -150 /
\plot  -87 -50
       -130 -75 /
\plot  -87 50
       -130 75 /

 \put {\Large $\bullet$} at -87 -03

\endpicture}
\mbox{
\beginpicture
\setcoordinatesystem units <0.08 pt,0.08 pt> \setplotarea x from -130 to 130, y from -180 to 180
\plotsymbolspacing=.2pt
\linethickness=2 pt
\setplotsymbol ({.})
\setlinear
\plot 0 100
      87 50 /
\plot  87 50
      87 -50 /
\plot  87 -50
       0 -100 /
\plot  0 -100
       -87 -50 /
\plot  -87 -50
       -87 50 /
\plot  -87 50
       0 100 /
\setsolid
\plot 0 100
      0 150 /
\plot  87 50
      130 75 /
\plot  87 -50
       130 -75 /
\plot  0 -100
       0 -150 /
\plot  -87 -50
       -130 -75 /
\plot  -87 50
       -130 75 /

 \put {\Large $\bullet$} at -87 -53
 \put {\Large $\bullet$} at -87  53

\endpicture} & None  \\ \hline
$G$ & $(\mathbf{v}\times \mathbf{\Sigma })(\mathbf{u}\times \mathbf{\Lambda }%
)$, $\mathbf{u},\mathbf{v}\in \mathbb{V}_{3}$ & $\frac{1}{2}(\mathbf{u}%
_{a}\cdot \mathbf{u}_{b})[3(\mathbf{n}\cdot \mathbf{v}_{a})(\mathbf{n}\cdot
\mathbf{v}_{b})-(\mathbf{v}_{a}\cdot \mathbf{v}_{b})]$ & e:%
\mbox{
\beginpicture

\setcoordinatesystem units <0.08 pt,0.08 pt> \setplotarea x from -130 to 130, y from -180 to 180

\plotsymbolspacing=.2pt
\linethickness=2 pt
\setplotsymbol ({.})
\setlinear
\plot 0 100
      87 50 /
\plot  87 50
      87 -50 /
\plot  87 -50
       0 -100 /
\plot  0 -100
       -87 -50 /
\plot  -87 -50
       -87 50 /
\plot  -87 50
       0 100 /
\setsolid
\plot 0 100
      0 150 /
\plot  87 50
      130 75 /
\plot  87 -50
       130 -75 /
\plot  0 -100
       0 -150 /
\plot  -87 -50
       -130 -75 /
\plot  -87 50
       -130 75 /

 \put {\Large $\bullet$} at -87 -03

\endpicture}
\mbox{
\beginpicture
\setcoordinatesystem units <0.08 pt,0.08 pt> \setplotarea x from -130 to 130, y from -180 to 180
\plotsymbolspacing=.2pt
\linethickness=2 pt
\setplotsymbol ({.})
\setlinear
\plot 0 100
      87 50 /
\plot  87 50
      87 -50 /
\plot  87 -50
       0 -100 /
\plot  0 -100
       -87 -50 /
\plot  -87 -50
       -87 50 /
\plot  -87 50
       0 100 /
\setsolid
\plot 0 100
      0 150 /
\plot  87 50
      130 75 /
\plot  87 -50
       130 -75 /
\plot  0 -100
       0 -150 /
\plot  -87 -50
       -130 -75 /
\plot  -87 50
       -130 75 /

 \put {\Large $\bullet$} at -87 -53
 \put {\Large $\bullet$} at -87  53

\endpicture} & None   \\ \hline
$E_{1}^{\prime \prime }$ & $%
\begin{array}{l}
s\Sigma _{x}(\bm\Lambda \cdot \hat{s}\mathbf{u})+\Sigma _{y}(\bm\Lambda
\times \hat{s}\mathbf{u}) \\
\hat{s}=\mathrm{diag}(1,s)%
\end{array}%
$, $%
\begin{array}{l}
\mathbf{u}\in \mathbb{V}_{3}, \\
s=\pm 1%
\end{array}%
$ & $\frac{2-s_{a}s_{b}}{2}\left[ (\mathbf{n}\cdot \mathbf{u}_{a})(\mathbf{n}%
\cdot \mathbf{u}_{b})-s_{a}s_{b}(\mathbf{n}\times \mathbf{u}_{a})(\mathbf{n}%
\times \mathbf{u}_{b})\right] $ & s:%
\mbox{
\beginpicture

\setcoordinatesystem units <0.08 pt,0.08 pt> \setplotarea x from -130 to 130, y from -180 to 180

\plotsymbolspacing=.2pt
\linethickness=2 pt
\setplotsymbol ({.})
\setlinear
\plot 0 100
      87 50 /
\plot  87 50
      87 -50 /
\plot  87 -50
       0 -100 /
\plot  0 -100
       -87 -50 /
\plot  -87 -50
       -87 50 /
\plot  -87 50
       0 100 /
\setsolid
\plot 0 100
      0 150 /
\plot  87 50
      130 75 /
\plot  87 -50
       130 -75 /
\plot  0 -100
       0 -150 /
\plot  -87 -50
       -130 -75 /
\plot  -87 50
       -130 75 /

 \put {\Large $\bullet$} at 0 90
\endpicture} & None   \\ \hline
\end{tabular}%
\end{table*}

The $\pi$-electron band in graphene is well described by the
closest-neighbor tight-binding model, $H_{0}=\gamma
_{0}\sum_{[\mathbf{r}_{A}\mathbf{r}_{B}]}\{c_{\mathbf{r}_{A}}^{\dagger }c_{
\mathbf{r}_{B}}+h.c.\}$ with hoping parameter $\gamma _{0}\sim 3\text{eV}$ \cite
{Review}. The sum runs over all pairs $[\mathbf{r}_{A}\mathbf{r}_{B}]$
of neighboring A and B sites of the lattice and $c^{\dagger }$/$
c$ are the on-site electron creation/annihilation operators
\cite{footnoteSpin}. An e-type
adatom attached to the bond between the cites
$\mathbf r_A'$ and $\mathbf r_B'$ creates a local perturbation
\begin{equation}
H_{a}=\gamma _{0}\xi _{1}(c_{\mathbf{r}_{A}^{\prime }}^{\dagger }c_{\mathbf{r
}_{A}^{\prime }}+c_{\mathbf{r}_{B}^{\prime }}^{\dagger }c_{\mathbf{r}
_{B}^{\prime }})+\gamma _{0}\xi _{2}(c_{\mathbf{r}_{A}^{\prime }}^{\dagger
}c_{\mathbf{r}_{B}^{\prime }}+c_{\mathbf{r}_{B}^{\prime }}^{\dagger }c_{
\mathbf{r}_{A}^{\prime }}),  \label{TBH}
\end{equation}
where $\xi _{1},\xi _{2}<1$
determine how the adatom affects the on-cite potential ($\xi_1$)
and the electron hopping amplitude between the cites ($\xi_2$).

The long-range RKKY interaction between two adatoms is due to the
perturbation of the electron spectrum near the Fermi energy and is
adequately described in terms of the four-component field,
 $\psi (\mathbf{r})=(\psi _{1},\psi _{2},\psi _{3},\psi _{4})^{T}$,
which is smooth on a scale of the lattice constant, $a$:
\begin{equation*}
c_{\mathbf{r}}=\left(\textstyle \frac{3}{4}\right) ^{\frac{1}{4}} a \times \left\{
\begin{array}{ll}
e^{i\mathbf{K}\mathbf{r}}\psi _{1}(\mathbf{r})+e^{-i\mathbf{K}\mathbf{r}%
}\psi _{4}(\mathbf{r}), & \quad \mathbf{r}=\mathbf{r}_{A} \\
e^{i\mathbf{K}\mathbf{r}}\psi _{2}(\mathbf{r})+e^{-i\mathbf{K}\mathbf{r}%
}\psi _{3}(\mathbf{r}), & \quad \mathbf{r}=\mathbf{r}_{B}%
\end{array}
\right. .
\end{equation*}
In the presence of an adatom $\psi(\mathbf r)$ obeys the
Hamiltonian \cite{Impurities}
\begin{align}
\hat{H}& =
v\int \psi ^{\dagger }(\mathbf{\Sigma }\cdot \mathbf{\hat{p}})\psi
d^{2}r+\hbar va\psi ^{\dagger }(\mathbf{r}_{a})\hat{W}_{a}\psi (\mathbf{r}_{a}) ,
\label{lambdaTa} \\
\hat{W}_{a} & =\lambda _{A_{1}}+\lambda _{E_{1}^{\prime }}\Sigma _{z}(
\mathbf{\Lambda }\cdot \mathbf{u}_{a})  \notag \\
& \quad+\lambda _{E_{2}}\Lambda _{z}(\mathbf{\Sigma }\cdot \mathbf{v}
_{a})+\lambda _{G}(\mathbf{\Lambda }\times \mathbf{u}_{a})\cdot (\mathbf{
\Sigma }\times \mathbf{v}_{a}),  \label{lambdaT}
\end{align}
with $v=\sqrt{3}a\gamma _{0}/2\hbar$
and $\Sigma _{x,y}=\sigma _{x,y}^{\mathrm{s}}\otimes \sigma _{z}^{\mathrm{v}}.$
The Pauli matrices $\sigma _{x}^{\alpha },\sigma
_{y}^{\alpha }$ and $\sigma _{z}^{\alpha }$ operate on the valley ($\alpha =
\mathrm{v}$) or sublattice ($\alpha =\mathrm{s}$) indices.
Together with $\Sigma _{z}=\sigma _{z}^{\mathrm{
s}}\otimes \mathbf{1}^{\mathrm{v}}$, matrices $\Sigma _{x,y}$ form a
representation of the $\mathrm{SU}_{2}^{\Sigma }$ algebra.
The first term in $H$ determines the Dirac electronic spectrum. It
possesses a "flavour" $\mathrm{SU}
_{2}^{\Lambda }$ symmetry generated by the three matrices $\Lambda
_{x,y}=\sigma _{z}^{\mathrm{s}}\otimes \sigma _{x,y}^{\mathrm{v}}$ and $
\Lambda _{z}=\mathbf{1}^{\mathrm{s}}\otimes \sigma _{z}^{\mathrm{v}}$,
satisfying $[\Lambda _{i},\Sigma _{j}]=0$. This symmetry manifests
of the conservation of the electron's valley index. Matrices $\Sigma $,
$\Lambda $ and their products $\Lambda _{i}\Sigma _{j}$ can be
arranged into irreducible representations (irreps) of
the symmetry group $\mathcal{G}$ of the honeycomb lattice
\cite{Kechedzhi,Basko}, which includes lattice translations,
$C_{6v}$ rotations and mirror refections.
All operators $\Sigma _{i}$ and $\Lambda _{j}$
change signs upon time inversion, therefore only products
$\Lambda _{i}\Sigma _{j}$ are time-inversion-symmetric and can
appear in the scattering matrix \cite{footnoteT} $\hat W_a$ ,
Eq.~\eqref{lambdaTa}, describing static perturbations
\cite{DisorderedG,FO} created by adatoms.

For an adatom of a general symmetry type $\hat{W}_{a},$ can be expanded into orbits in the irreps
\cite{footnoteG} of $\mathcal{G},$ $\hat{W}_{a}=\sum
\lambda _{i}\mathcal{W}_{i}$.  The classification of
orbits by the irrep and the symmetry type of adatom
is given in the second column of Table \ref{tab:invars}.
In particular, the matrix $\hat{W}_{a}$ of an
e-type adatom (or the H-H dimer) is expanded into orbits in the
irreps $A_1, E_1', E_2$ and  $G,$ corresponding to the
four terms in Eq.~\eqref{lambdaT}.
Vectors $\mathbf{u}$ and $\mathbf{v}
$ in Eq. (\ref{lambdaT}) take values in the set $\mathbb{V}_{3}$ shown in
Fig.~\ref{Fig01}: three unit vectors on the x-y plane, at $120^{\circ }$
angles. Each given bond of graphene lattice is characterized by a pair $(
\mathbf{u},\mathbf{v)}$ as shown in Fig.~\ref{Fig01}. The distribution of $
\mathbf{u}$ and $\mathbf{v}$ forms a periodic pattern with three times the
graphene lattice period: the intervalley scattering implies the
momentum transfer $\Delta \mathbf{K}=\mathbf{K}-\mathbf{K}^{\prime }$ such
that $3\Delta \mathbf{K}$ is a reciprocal lattice vector. Thus, $ \hat{W}_{a}$
in the e-type case is periodic on a
superlattice, whose unit cell is three times as big as
graphene's and contains nine distinguishable C-C bonds. The three
non-trivial terms in Eq. (\ref{lambdaT}) resemble the
coupling of an electron to the in-plane $\Gamma $- and K-point
phonons \cite{Basko}. Indeed, the $E_{1}^{\prime }$ term parameterized by
the vector $\mathbf{u}$ resembles the effect of the
K-point breathing phonon mode. Below, the ensemble average of $\mathbf{u}$
will play the role of the order parameter.
The $G$ term is similar to
the 4-fold degenerate K-point lattice mode. The $E_{2}$ term resembles
the uniaxial strain due to a
$\Gamma $-point optical phonon. The
parameters $\lambda _{i}$ in $\hat{W}_{a}$ are specific for particular atoms. For the model in Eq.~\eqref{TBH}, $\lambda
_{E_{1}^{\prime }}=\lambda _{E_{2}}=\xi _{2}$ and $\lambda _{A_{1}}=\lambda
_{G}=\xi _{1}$.

Unlike e-type adatoms,  the perturbation introduced
by an s-type adatom (\textit{e.g.}, H on a lattice site) cannot be
related to the in-plane phonons, since it explicitly
distinguishes A and B sublattices of graphene. The two orbits encountered
in the matrix $\hat{W}_{a}$ in this case can be related to the
out-of-plane phonons in the presence
of a transverse electric field ($z\rightarrow -z$ asymmetry): $B_{1}$
resembling the $\Gamma $-point and $E_{1}^{\prime \prime }$ the K-point
phonon. In both cases, the A/B-residency of an adatom is accounted for by $
s=\pm 1$.

Each type of the adsorbents listed in Table \ref{tab:invars} creates Friedel
oscillations of the electron density breaking the symmetry in the
same way as the adatom does. The polarization caused by one
adatom extends over long distances, thus leading to interaction between
the adatoms. The energy of this interaction between a pair of
adatoms at a distance $|\mathbf{r}_{ab}|\gg a$ can be expressed through
the imaginary time (Matsubara) Green function of electrons in a clean graphene,
$G(\mathbf{r},\tau )$ \cite{footnoteGmom} as
\begin{align}
F&_{ab}=2\hbar v^2a^2\mathrm{Tr}\int_{-\infty }^{\infty }d\tau \hat{W}_{a}
\hat{G}(\mathbf{r}_{ab},\tau )\hat{W}_{b}\hat{G}(-\mathbf{r}
_{ab},-\tau ),  \notag \\
\hat{G}&(\mathbf{r},\tau )=-\frac{1}{4\pi }\frac{v\tau +i\mathbf{\Sigma }
\cdot \mathbf{r}}{(v^{2}\tau ^{2}+r^{2})^{3/2}}.
\label{Fab}
\end{align}
Here, trace
is taken over the valley and sublattice indices, and spin degeneracy is
taken into account. Equation~\eqref{Fab} yields a long-range pair
correlation energy,
\begin{eqnarray}
F_{ab} &=&-\frac{\hbar va^{2}}{4\pi }\frac{\Omega _{ab}}{r_{ab}^{3}},\qquad
\mathbf{n}=\frac{\mathbf{r}_{ab}}{r_{ab}},  \notag \\
\Omega _{ab} &=&\frac{1}{16}\mathrm{\ Tr}\left[ \;\hat{W}_{b}\hat{W}_{a}-3%
\hat{W}_{a}(\mathbf{\Sigma }\cdot \mathbf{n})\hat{W}_{b}(\mathbf{\Sigma }%
\cdot \mathbf{n})\right] .  \label{Omint}
\end{eqnarray}

Interaction between  adatoms of the same type depends on both the
type of the adatoms and the parameters $\lambda _{i}$ characterizing the
adatom-electron coupling in each symmetry-breaking interaction channel $
\mathcal{W}_{i}$:
\begin{equation}
\Omega _{ab}=\sum_{i}\lambda _{i}^{2}\Omega _{ab}^{(i)}.
\label{Decomposition}
\end{equation}
For the model in Eq.~\eqref{lambdaT}
with $\lambda _{E_{1}^{\prime }}$ larger than other coupling
parameters, $\Omega _{ab}\approx \lambda _{E_{1}^{\prime }}^{2}\mathbf{u}
_{a}\cdot \mathbf{u}_{b}$, so that the interaction between lateral degrees
of freedom of the two adsorbents looks like an isotropic ferromagnetic
exchange. Therefore the adatoms tend to occupy preferably bonds with the same
$\mathbf{u}$ and form a partially ordered state shown in Fig.~\ref
{Fig01}. The transition to such a state is described by a special case of
random
exchange three-state Potts model
\cite{Wu}:  'spins' $\mathbf{u}$ reside on randomly distributed with
density $\rho$ sites and experience pairwise exchange interaction
$- J \mathbf u_a\cdot \mathbf u_b/r^3.$
According to recent cluster Monte-Carlo\cite{Us}
studies this model undergoes an order-disorder
transition at a critical temperature $T_c\approx 8 \rho^{3/2}J.$ For the
Eq.~\eqref{TBH} model $T_c$ is evaluated as
\begin{equation}
T_{c}\approx 0.6\lambda _{E_{1}^{\prime }}^{2}(a^{2}\rho )^{3/2}
\frac{{\hbar v}}{a}.  \label{Tc}
\end{equation}
\begin{figure}
\includegraphics[width= 0.48 \textwidth]{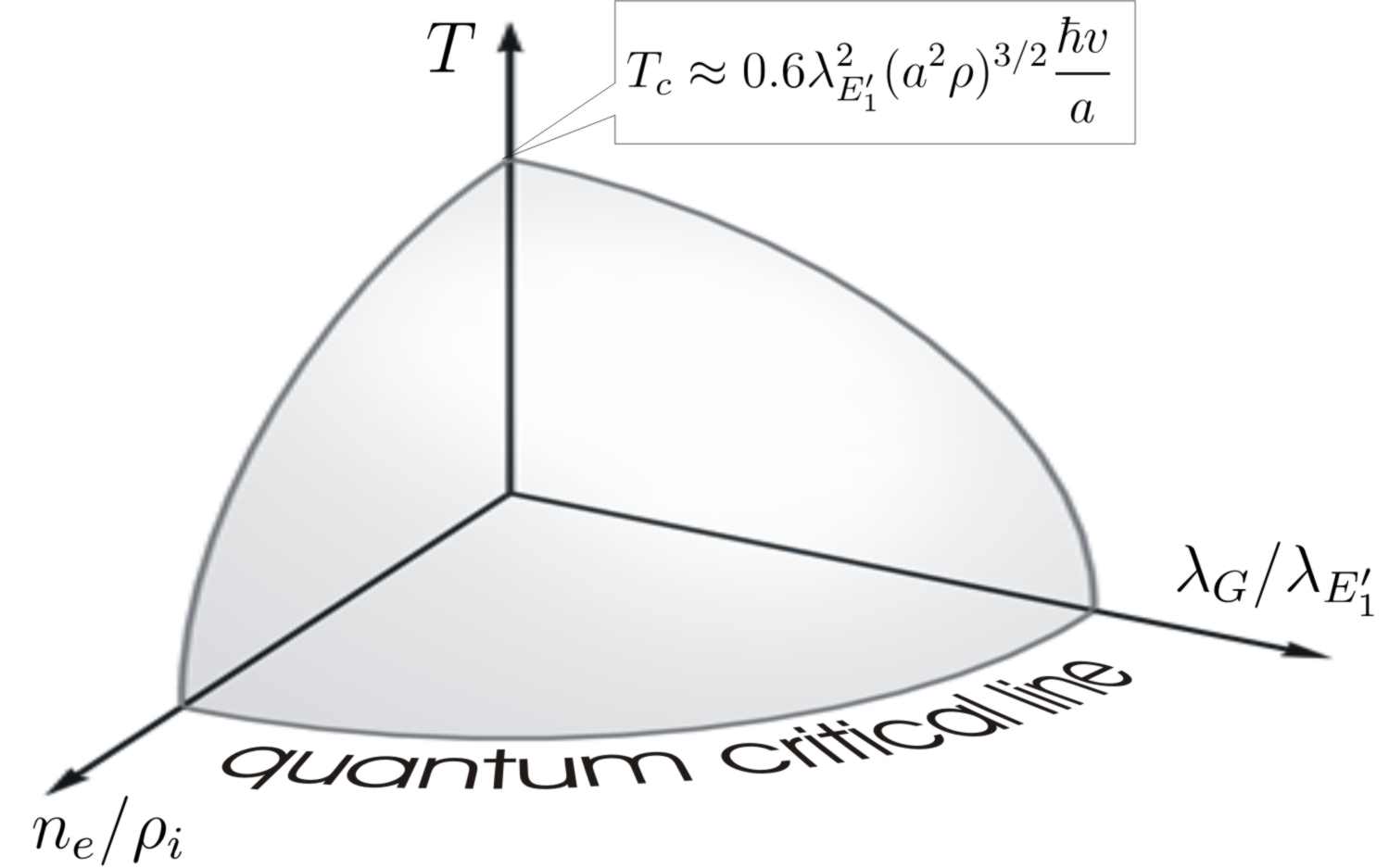}
\caption{The schematic phase diagram of the epoxy-bonded adatom system on
graphene. The critical surface separates the gapped ordered phase (Fig.~\ref{Fig01})
from the disordered phase. }
\label{Fig2}
\end{figure}

Consider now effects of other terms in \eqref{Decomposition}
allowed by the e-type symmetry.
These terms are listed in the rows $A_1$, $E_2$ and $G$
in Table \ref{tab:invars}. The symmetric perturbation
parameterized by $\lambda _{A_{1}}$ leads to the repulsion between
adatoms regardless of which bonds of the extended
supercell they occupy. The coupling parameterized by $\lambda _{E_{2}}$ causes
an anisotropic 'antiferromagnetic' interaction of the alternative set of
Potts "spins"  $\mathbf{v}_{a}$. Frustration precludes ordering of
$\mathbf{v}_{a},$ which could lead to a
unilateral deformation of the lattice. However,
a presence of $\lambda _{E_{2}}^{2}\Omega _{ab}^{(E_{2})}$
in $\Omega _{ab},$ Eq. (\ref{Decomposition}) does
not affect the ordering of vectors $\mathbf u_a,$
at least up to the quadratic order in $\lambda
_{E_{2}}.$
The isotropic interaction leading
to the ordering only competes with the anisotropic interaction between
adatoms caused by the last term in $\hat{W}_{a}$, Eq.~(\ref{lambdaT})
(the fifth row of Table \ref{tab:invars}). If the latter is strong, $
\lambda _{G}>\lambda _{E_{1}^{\prime }}$, it suppresses ordering by
frustration. As a result, $T_c$ of the order-disorder
transition decreases with increasing ratio
$\lambda _{G}/\lambda _{E_{1}^{\prime }}$
until it vanishes at a quantum critical point.

Increasing the density $n_e$ of mobile carriers in graphene should
also
suppress $T_c.$  Indeed, at finite $n_e$
the RKKY interaction develops Friedel oscillations\cite{FO},
$F\propto (\mathbf{u}_{a}\cdot \mathbf{u}_{b})
\sin ^{2}(k_{F}r)/r^{2},$ which lead to a random sign
of the exchange coupling between adatoms at distance
$r_{ab}>\sqrt{1/n_e}.$ At sufficiently large ratio $n_e/\rho$
this effect should completely destroy ordering of Potts "spins".
These considerations are illustrated by the phase diagram in Fig.~\ref
{Fig2}, where the "quantum critical line" corresponds to
the parametric
condition $T_c(n_e/\rho, \lambda_G/\lambda_{E'})\to 0.$
Further analysis of these transitions is beyond the scope of this paper.

The interaction of s-type adsorbents residing on the honeycomb lattice sites
(\textit{e.g.}, hydrogen) is described in rows 2 and 6 of Table~\ref
{tab:invars}. Each s-type adatom can be characterized by the
Ising "spin" $s_a$ taking the value $+1$ or $-1$ depending on which sublattice
A or B it occupies. These spins
may, potentially, establish sublattice ordering. However, there
will be no ordering of the
"spins" $\mathbf u_a:$  the interaction
described in Table~\ref{tab:invars} is anisotropic and causes
frustrations. In contrast, a pair of hydrogens forming an H-H
dimer on the nearest A/B sites \cite{H2-on-G} falls into the same symmetry
class as e-type adatoms and can establish the same type of ordering. It has been
noticed that in hydrogenated graphite both configurations of H atoms are
present \cite{H2-on-G}. Since the decomposition of the interaction in
Eq.~\eqref{Omint}
suggests the absence of mutual correlations between adsorbents
of the e-type and s-type, we expect ordering of the H-H dimers on
graphene, even if only a fraction of hydrogens covering
the flake is dimerised.

The predicted ordering will strongly influence transport and
optical properties of the material. As temperature approaches $T_{c}$ from
above large clusters of ordered phase will act as intervalley-scattering
Bragg mirrors for electrons. Back-scattering from such mirrors should
lead to a power-law increase of
resistivity near $T_{c}$ \cite{TBP}. At $T<T_{c}$, the spectrum of electrons
becomes gapful,
\begin{equation}
\varepsilon =\pm \sqrt{v^{2}p^{2}+\Delta ^{2}},\;\;\Delta =\frac{
\hbar v}{a}\lambda _{E_{1}^{\prime }}\rho a^{2}\gg T_{c},
\end{equation}
as can be seen from the mean-field Hamiltonian \cite{Kekule1},
$
\langle \hat{H}\rangle =v\mathbf{\Sigma }\cdot \mathbf{p}+
\hbar v\rho a\lambda _{E_{1}^{\prime }}\Sigma _{z}\mathbf{
\Lambda }\cdot \mathbf{u,}
$
valid for $p\ll \hbar \rho ^{1/2}$.
At low carrier density this will lead to the activated
transport regime typical of semiconductors.

Partial ordering of e-type adsorbents should also be manifest in
the structure of the D-peak in the Raman spectrum.
 The D-peak is associated with the excitation
of one optical phonon at momentum $\mathbf K$ (or $\mathbf K'$)
and is forbidden by momentum conservation  in pristine graphene.
In the presence of random scatterers, the D-peak is seen as a
low-intensity, $I(T>T_{c})\propto \rho \lambda _{E_{1}^{\prime }}^{2},$
feature strongly broadened due to the disorder-induced uncertainty in
the emitted phonon momentum. Domains of ordered adsorbent,
with the size $L$, will scatter electrons between valleys $K$ and $
K^{\prime }$ coherently. This will enhance the intensity of the otherwise
forbidden transition,\ $I(T<T_{c})\propto \rho \lambda _{E_{1}^{\prime
}}^{2}\times \rho L^{2}\gg g(T>T_{c})$, and restrict the uncertainty of
the
emitted phonon momentum to $\delta q\sim 1/L$. Thus, one can
predict that the ordering of adatoms
abruptly enhances and narrows  the
D-peak in the Raman spectrum. To mention, an
observation of a hopping conductivity accompanied by
a sharp
high-intensity D-line in the Raman spectrum in graphene exposed for a long
time to hydrogen atmosphere has been reported in Ref. \cite{Geim2}.
We predict a similar behavior of the ARPES spectrum of graphene: ordering
should strongly enhance the photoemission of electrons from the center of the Brillouin zone
at energies close to the Fermi energy.

The work was supported by the Lancaster-EPSRC Portfolio Partnership, ESF CRP
SpiCo, and US DOE contract No. DE-AC02-06CH11357.


\begin{thebibliography}{99}
\bibitem{footnoteSpin}
In this Letter we are focusing on interactions conserving spin and suppress
spin indices throughout.
\bibitem{footnoteT} Strictly speaking, the second term in the Hamiltonian,
Eq.~\eqref{lambdaT} is not well defined due to ultraviolet problems and is
written in this form for illustrative purposes. In a more rigorous approach
$\hat W_a$ should be interpreted as a scattering $T$-matrix
defining the long-distance asymptotic form of the electron wave scattered off the defect.
\bibitem{footnoteG} The scattering matrix $\hat W_a$ transforms as a tensor
under the symmetry group $\mathcal G$ of the lattice and can therefore be decomposed
into projections onto irreducible spaces
$\hat W_a=\sum_i \mathcal W_i$ . For an adatom preserving some subgroup of
a symmetry group of the lattice each $\mathcal W_i$ is an invariant under
this subgroup. As the position of the adatom in the lattice
is changed by applying elements of $\mathcal G,$
each $\mathcal W_i$ sweeps an orbit of $\mathcal G$ in the corresponding
irreducible representation.
\end{thebibliography}
\end{document}